# On the Experimental Evidence for Possible Superconductivity in LK 99

H. Singh[1]*, A. Gautam[1]*, M. Singh[2]*, P. Saha[2]*, P. Kumar[2], P. Das[2], M. Lamba[2], K. Yadav[2], P. K. Mishra[1] and S. Patnaik[2] and A. K. Ganguli[1,3]

[1]*Department of Chemistry, Indian Institute of Technology, New Delhi 110016, India*
[2]*School of Physical Sciences, Jawaharlal Nehru University, New Delhi 110067, India*
[3]*Indian Institute of Science Education and Research, Berhampur, Odisha 760010, India*

[a)]Corresponding author: ashok@chemistry.iitd.ac.in, spatnaik@jnu.ac.in
\* These authors have made main contributions

**Abstract.** The desire to create an energy efficient world is bound to be incomplete without the discovery of a room temperature superconductor at ambient pressure. A recent report on the room-temperature ambient-pressure superconductor has inspired scientists to study the Cu doped Lead apatite named as LK-99. Here, we have synthesized Cu doped LK-99 and Ni-doped LK-99 compounds and studied their temperature dependent transport and magnetization behavior. Inspite of the presence of impurity phase $Cu_2S$, the temperature dependent resistance shows an insulating nature of the sample. The radio frequency penetration depth measurement unveils the absence of diamagnetic flux expulsion in this sample. The temperature dependent ac susceptibility measurements reveal the paramagnetic nature of the Ni doped LK-99.

## INTRODUCTION

There are times in our scientific lives when appearance of USOs (Unidentified Superconducting Objects) become much more frequent than sightings of UFOs. One such occasion has recently arisen with the report of the room-temperature ambient-pressure superconductor LK-99[1,2]. Over the last two weeks, it has expeditiously motivated a massive surge of interest in synthesizing this Cu doped Lead apatite compound. The exhibition of superconductivity in this material is evidenced not only through electrical transport but also by magnetic measurements [1]. More intriguingly, the levitation of the as grown material over the magnet at room temperature has led to an outburst of worldwide exuberance.

But one has to be careful to read the experimental data with regard to genuine characterization of the superconducting state. In Wikipedia and elsewhere, a superconducting state is commonly ascribed to a zero electrical resistance state with concurrent onset of diamagnetism. However, experimentally it is impossible to ascertain zero electrical resistance state. The "zero" voltage measurements are always limited by resolution of voltmeters. Further, in the nascent stage of processing of new superconductors, the polycrystallinity of samples leads to percolative current paths with negligible inter-grain connectivity. So the generally accepted criteria for observation of superconductivity is the measurement of electric field less than 1μV/cm across the voltage taps. And with the application of magnetic field one would expect the offset of 1 μV/cm criteria to move to lower temperature. Similarly, there is no material which is not diamagnetic; only when other possible magnetic correlations become quiet, diamagnetic susceptibilities come to the fore. The most dependable magnetic signature of superconductivity therefore is splitting of zero field cooled and field cooled magnetization as a function of temperature. The third convincing criteria is the I-V characteristic where on reaching the critical current density, one would expect onset of normal state resistivity. We need to remember that many structural/ordering transitions can lead to rapid change in resistance with simultaneous variation in magnetic properties. These will not have anything to do with formation of a superconducting condensate. Case in point is the impurity phase of Copper sulfide in LK99. It is well known that $Cu_2S$ undergoes a structural phase transition at ~104 C accompanied by sharp changes in resistivity and magnetization properties. The question is the following: can a ~5% impurity phase account for drastic change in diamagnetic property along with flux trapping?

The theory offered by lead authors of LK99 rests on the fact that substitution of Cu at Pb sites of Lead apatite leads to a shrinkage in the unit cell volume resulting in a induced superconductivity in LK-99 [3]. This is reminiscent of the key idea with regard to three time increase in transition temperature between LSCO cuprate and YBCO cuprate. In a matter of few days many reports have attempted to synthesize the same sample and test the reproducibility of the results [4,5]. Till now, all the experimental reports have negated the claims of superconductivity in this compound through basic transport and magnetic measurements [4,5,7-9]. However, a recent report by Q Hou et al [10] revealed the presence of superconductivity in LK-99 below 110K which is way lower that the transition temperature obtained by Lee [1]. But there is no evidence of 1 µV/cm criteria in this paper. Another report has shown the magnetic levitation of LK-99 over a magnet with transition temperature to be around ~325K [11]. Lee et al., have suggested two possible mechanisms for the induction of superconductivity in this compound: one is the superconducting quantum well and the other is the strong correlation due to the enhancement in coloumb interaction achieved by the doping of Copper [1,2]. Theoretical studies have reported the formation of flat bands near the fermi level due to the doping of Cu in LK-99 [12-15]. A recent density functional calculation has claimed the presence of bulk superconductivity in LK-99 only if the Cu can be substituted to the appropriate Pb (1) sites (one which forms a network comprising $PbO_6$ prisms that are corner shared with $PO_4$ tetrahedra) [6]. Hence the above results hint towards the challenge in synthesis of LK-99 where Cu occupies the Pb (1) sites that may provide some insight towards the absence of superconductivity in previous reports [4,5].

Here, we have synthesized LK-99 and Ni doped LK-99 and have conducted the temperature dependent resistance and magnetization and penetration depth measurements. Our initial effort shows that the samples are mired by impurity phase $Cu_2S$, but even then we find no evidence of superconductivity in LK-99.

## EXPERIMENTAL TECHNIQUES

The synthesis of LK-99 and Ni doped LK-99 involves four steps [1]. Firstly, the synthesis of $Pb_2SO_5$ (lanarkite) requires two essential components i.e; $PbSO_4$ and PbO. The synthesis of $PbSO_4$ can be accomplished through the subsequent chemical reaction:

$$Pb(NO_3)_2 \;+\; H_2SO_4 \;\rightarrow\; PbSO_4$$

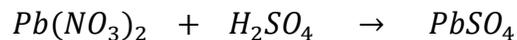

After that, PbO and PbSO4 powders were thoroughly combined in a 1:1 molar ratio using an agate mortar and pestle. Subsequently, the resulting mixture was placed into an alumina crucible and subjected to a 24-hour reaction at 725 °C within a furnace. Once the reaction concluded, a white sample was acquired, which was then finely ground using the mortar.

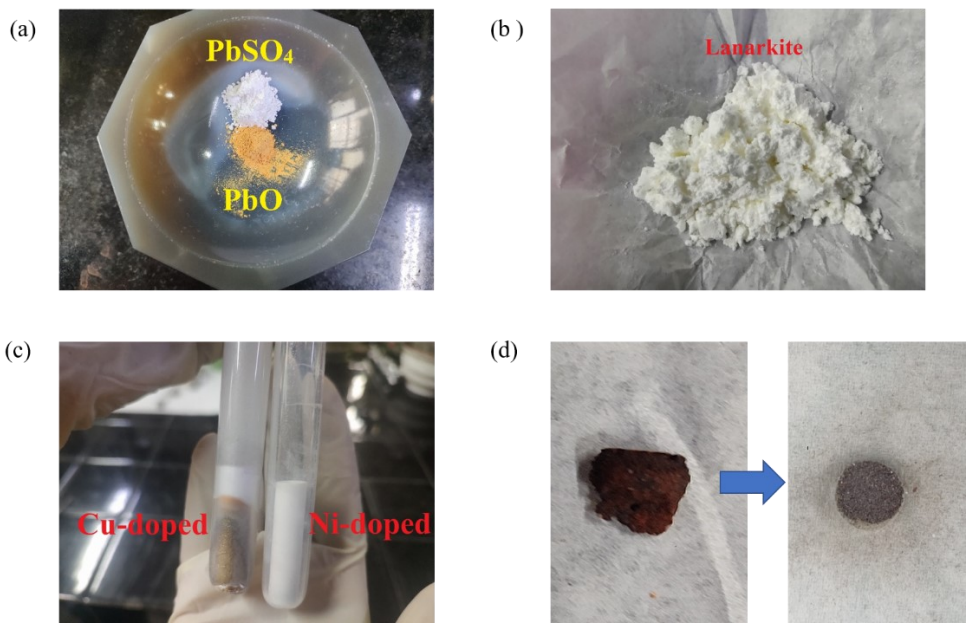

FIGURE 1 (a) shows the powders of PbSO4 and PbO (b) shows the image of the Lanarkite obtained after the reaction. (c) shows the image of Cu and Ni doped LK-99 after taking out of the furnace. (d) The first part of the figure shows the extracted Copper doped Lead Apatite from the sealed tube and the brick red colour of the compound indicates unreacted copper. The second part shows the grey colored sample obtained after grinding in mortar pestle.

For the synthesis of $Cu_3P$ Copper (Cu) and phosphorus (P) powders were combined in 3:1 composition ratio. The resulting mixture was then placed inside a quartz tube, which was subsequently sealed under a vacuum of $10^{-5}$ Torr. This sealed tube was subjected to a 48-hour reaction at 550 °C within a furnace. The synthesis process was similarly applied to produce $Ni_3P$. Standard solid-state reaction approach was employed for the synthesis of Cu doped apatite and Ni-doped apatite. Lanarkite and $Cu_3P/NI_3P$ were thoroughly blended in a 1:1 molar ratio using an agate mortar and pestle. The resulting sample was then placed within a reaction tube, which was sealed under a vacuum of $10^{-5}$ Torr. This assembly underwent a reaction process at a temperature of 925°C for a duration of 24 hours. Fig.1(c and d) shows the final products formed after the above reaction process.

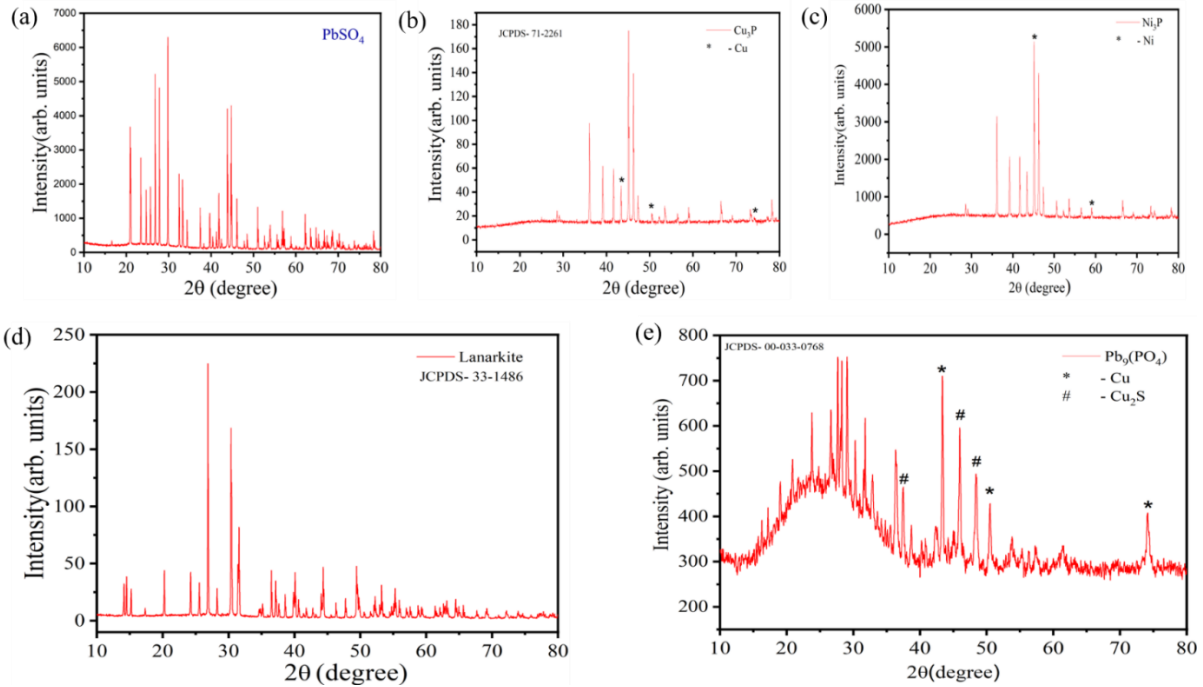

FIGURE 2. (a) shows the PXRD of PbSO4. (b) shows the XRD pattern of Cu3P. (c) shows the XRD pattern Ni3P. (d) shows the XRD pattern of Lanarkite (e) shows the XRD pattern of Cu doped Lanarkite.

## RESULTS AND DISCUSSION

### Structural Characterization

The phase formation of starting precursors ($PbSO_4$, $Cu_3P$, $Ni_3P$, $Pb_2(SO_4)O$ and the final product (Cu and Ni-doped Lead Apatite) was verified by room-temperature powder X-ray diffraction technique using Bruker D8 Advance diffractometer. The PXRD has been shown in fig.2. The diffraction peaks of $PbSO_4$ shown in fig. 2(a) were matched using the JCPDS file no- 71-2261. The PXRD pattern exhibits no peaks of impurity phases.

The diffraction peaks of $Cu_3P$ is shown in fig. 2(b). The pattern were matched using the JCPDS file no- 71-2261. The diffraction pattern also consists of diffraction peaks from unreacted copper at 2θ corresponding to 43.45, 50.5 and 74.21 degrees. The diffraction peaks of $Ni_3P$ (fig. 2(c)) were matched using the JCPDS file no- 71-2261. The observed diffraction pattern exhibit peaks of unreacted Nickel. The diffraction peaks of Lanarkite (fig. 2(d)) were matched using the JCPDS file no- 33-1486. The observed diffraction pattern exhibits no signature peaks of impurity phases. The powder X-ray diffraction pattern of Cu doped Lanarkite was observed using Bruker D8 Advance diffractometer. The PXRD pattern displays the presence of $Pb_9(PO_4)$ matched using the JCPDS file no. 00-033-0768. In addition to this, impurity peaks of Cu and $Cu_2S$ is also observed which is possibly appearing due to the reaction of unreacted copper with sulfur.

**Transport and Magnetization Measurements**

The room temperature resistance of the grown LK-99 was around 1kΩ. Fig 3(a) shows the temperature dependent resistivity for LK-99 for a temperature range of 150K to 200K. As depicted in the fig. 3a the resistivity shows an exponential decay with temperature. The same trend of behavior is continued for the high temperature range as well in fig. 3(b). This behavior implies the insulating nature of the sample. Fig 3(c) shows the penetration depth measurement using the tunnel diode oscillator in the temperature range of 250K to 300K. This data illustrates the absence of any significant change in the frequency which clearly rules out the presence of diamagnetic flux expulsion in this compound. Fig 3(d) shows the temperature dependent ac susceptibility of Ni-doped LK-99. This behavior has been plotted with Gadolinium which is ferromagnetic above 293K. As can be seen the Ni doped LK-99 shows no considerable change with temperature. After subtracting the background, we obtain a linear decrease in susceptibility with increasing temperature. This reveals the paramagnetic nature of this compound.

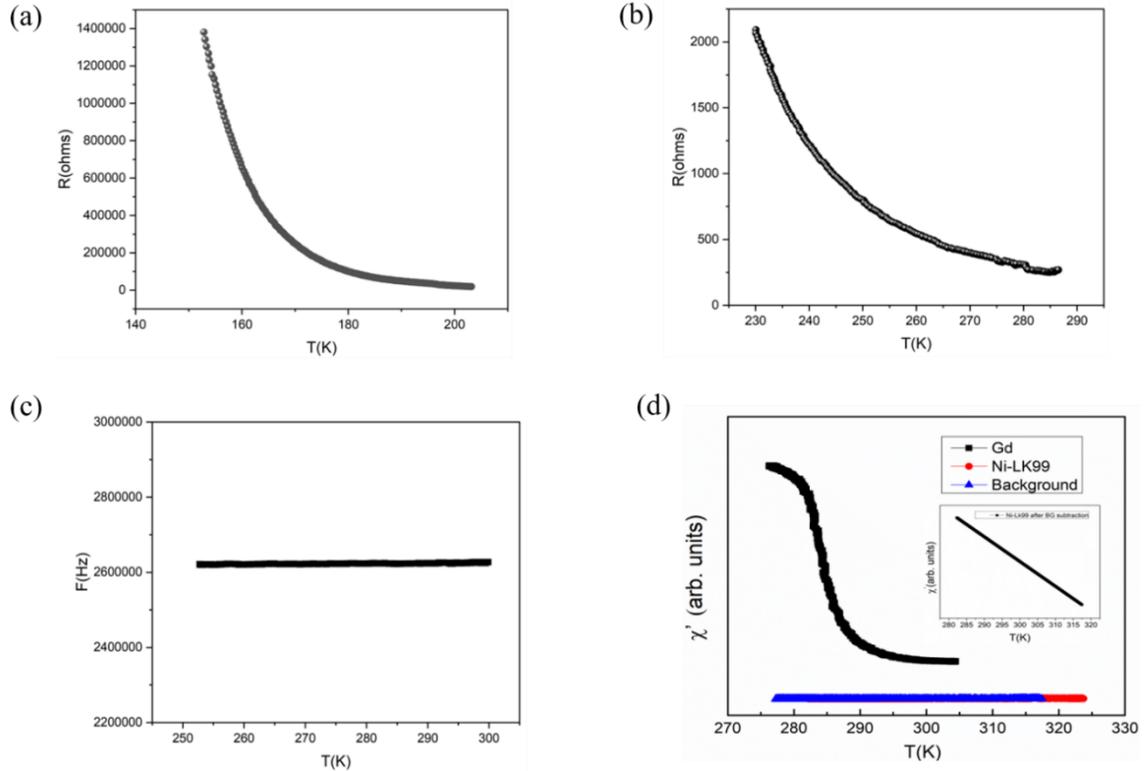

FIGURE 3. (a) shows the temperature dependent resistivity in the low temperature region. (b) shows the change in resistivity in the high temperature region. (c) shows the temperature dependent penetration depth measurement for LK-99. (d) shows the temperature dependent ac susceptibility measurement for Ni-doped LK-99.

## CONCLUSION

In summary, we have synthesized Cu and Ni doped Lead apatite. Temperature dependent resistivity reveals insulating nature of LK-99 for low and high temperature ranges. Temperature dependent penetration depth measurement rules out any signature of diamagnetic flux expulsion in this sample. Temperature dependent ac susceptibility measurements reveal the paramagnetic nature of Ni doped LK-99. Our results do not support any observation of superconductivity in LK99 although $Cu_2S$ is present as an impurity phase.